\documentclass[12pt]{article}
\usepackage{bm}
\usepackage{graphicx}

\title{QCD AGAINST BLACK HOLES$^*$ ?}
 
\vspace{5mm}

\author{Ilya I. Royzen}
\date{}
\begin{document}
\maketitle \centerline{\em{P.N. Lebedev Physical Institute of RAS}}

\centerline{e-mail: $<royzen@lpi.ru>$}
\addtolength{\baselineskip}{6pt}
\vspace*{20mm}
\begin{abstract}



 Along with compacting baryon (neutron) spacing, two very important factors
come into play at once: the lack of self-stabilization within a compact 
neutron star (NS) associated with possible black hole (BH) horizon appearance 
and the phase transition - color deconfinement and QCD-vacuum reconstruction - 
within the nuclear matter. That is why both phenomena should be taken into
account side by side, as the gravitational collapse is 
considered. Since, under the above transition, the hadronic-phase vacuum
(filled up with gluon and chiral $q\bar q$-condensates) turns into the "empty"
(perturbation) subhadronic-phase one and, thus, the corresponding (very high)  
pressure falls down rather abruptly, the formerly cold (degenerated) nuclear 
medium starts to implode into the new vacuum. If the mass of a star is 
sufficiently large, then this implosion produces an enormous heating, which 
stops only after quark-gluon plasma of a temperature about 100 MeV (or even 
higher) is formed to withstand the gravitational compression (whereas the 
highest temperatures of supernovae bursts are, at least, one order lower). As
a consequence, a "burning wall" must be, most probably, erected on the way
of further collapsing the matter towards a black hole formation.

\end{abstract}

\vspace*{3.5cm}

-----------------------------------------------------------------------

$^*$Talk at the 4th International Sakharov Conference, 
Moscow, May 18-23, 2009.

\newpage
\section{Twofold signal of neutron star instability}

Two mechanisms underlying the neutron star instability are to be discussed 
below: the first one consists in hadronic phase $\longrightarrow$ 
subhadronic phase (HPh $\to$ SHPh) transition within nuclear matter (it is 
described here in more detail) and the second one, which is rather familiar, is 
shutting to BH. They are engaged in "competition" with each other, however they
make the star to evolve in absolutely alternative ways; 
thus, the  main point is to understand, which one comes before into operation
\footnote{Here the non-rotating objects are under discussion only. Allowing for 
rotation would undoubtedly lead to the enhancement of instability}.

\subsection{Phase transition in nuclear medium}

Schematically, this transition is depicted as follows: \\

\qquad\qquad\qquad \,\, {QCD HPh  \qquad\qquad\qquad  
{$\Longleftrightarrow$} \qquad\qquad\quad QCD SHPh}\\ 

\hspace*{3.5cm} {$\Downarrow$ \hspace*{7cm}  $\Downarrow$ }\\

\hspace*{0.0cm} {$P^0_{vac}=-\varepsilon^0_{vac}\simeq
\varepsilon_{n}\simeq 
5\,10^{-3}$ {GeV$^4$} \quad\qquad $\Longleftrightarrow$ \quad\qquad $P_{vac}=
-\varepsilon_{vac}\rightarrow\,0$}\\

\hspace*{3.5cm} {$\Downarrow$ \hspace*{7cm}  $\Downarrow$ }\\

\hspace*{0.0cm}{$P_{tot}\simeq\,P^0_{vac}\,\,
\lbrack rarefied\,\,gas\,\,of\,\,hadrons\rbrack$ \quad\quad 
$\Longleftrightarrow$ \quad\quad {$P_{tot}=P_{vac}\,\,+\,\,P_{part}$} \\

\noindent Here ($\varepsilon^0_{vac}$, $P^0_{vac}$) and ($\varepsilon_{vac}$, $P_{vac}$)
stand for the vacuum (energy, pressure) in HPh and SHPh, respectively, while $P_{part}$
is the pressure of particles and $P_{tot}$ is the overall pressure within the nuclear
medium.
  
One has to consider two conceivable scenarios of this phase transition 
\cite{R_2008,R_2009,R-F-Ch} - the
hard scenario, when the HPh transforms at some density (pressure) directly 
(stepwise) into the current quark state (this is a "conventional" phase 
transition), and the soft one, which admits an intermediate state in between.
This state is attributed with deconfined 
dynamical quarks (valons) - quazi-particles of non-fixed mass, which 
diminishes along with the density (pressure) increase.
It is shown below that both scenarios result in developing strong 
instability under the phase transformation.

\newpage
{\bf\boldmath 1.\quad$Hard\,\,(stepwise)\,\,scenario:$ \quad SHPh is just 
$P_{vac}\,=\,\varepsilon_{vac}\,\equiv\,0$}\\ \\
This implies that the chiral symmetry restores and the current quarks - 
almost massless ($u,d$)- \,and \,$\sim$150-MeV $s$-ones - are emerged 
promptly, as neutrons crush down. It is illustrated in Fig.1 \cite{R_2008} that 
transition into degenerate ("cold") quark gas is ruled out: this scenario should 
unavoidably result in immediate development 
of a collapse into the new ("empty") vacuum and, thus,  
in an enormous heating \footnote{Two pressures - HPh-vacuum and SHPh-particle ones -
become equal only when the particle number density is 3-4 times higher (point {\it B} in
Fig.1). It is worth noting that the neutrinos get 
stuck under relevant densities and, thus, there is no way for an "instant" 
energy release.} (see an estimate below) of the nuclear medium at 
the phase transition point.  

\begin{figure}[h]
\includegraphics[width=10cm]{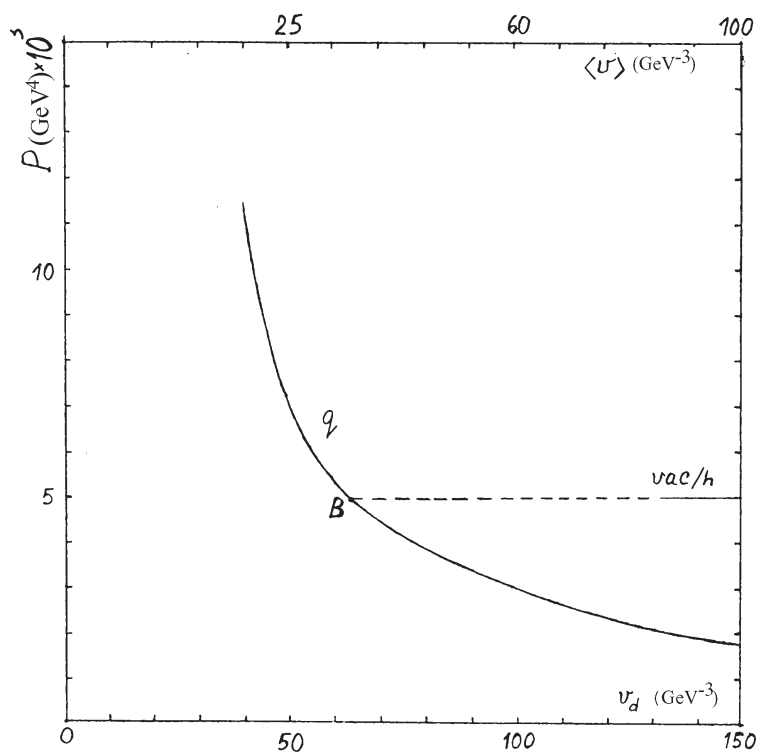}
\caption{The pressure of non-perturbation QCD vacuum
  condensate in the HPh (horizontal segment $vac/h$) vs the pressure of   
  degenerate (''cold'') perfect gas of $(u,d,s)$ current quarks 
  (curve $q$). 
  As the particle density approaches the critical value (neutron spacing becomes 
  compact, particle specific volume is
  $\langle v\rangle\,\simeq$ 100 GeV$^{-3}$), the occurence of a giant gap 
  between the HPh- and SHPh-phase pressures is quite well pronounced  - the former 
  is about three times as large as the latter one.  }
\end{figure}

\newpage

2.\quad{\bf\boldmath $Soft\,\,scenario$: \quad No stepwise HPh 
$\longleftrightarrow$ SHPh transition (crossover)}\\
 
In other words, as the neutrons "get in touch with each other" and loss their
identity, the degrees of freedom which come into life are associated with some 
hypothetical quazi-particles - massive dynamical quarks (valons) 
\cite{Sh,F,F-Ch,Cley,AKM}: they become the first color objects to be unleashed (color 
deconfinement); then, both the valon masses and vacuum condensate pressure
decrease along with the particle density increase \cite{R_2009}; finally, the 
valons turn into the current quarks and the chiral symmetry restores. Thus,
$P_{vac}\,=\,-\varepsilon_{vac}\,\to\,0$ more or less gradually. 

A reasonable approach \cite{R_2009}, which describes the degenerate valonic gas at 
particle energy 
densities $\varepsilon\,\geq\,|\varepsilon^0_{vac}|$, is based on the EoS:

\begin{equation} 
\varepsilon\,=
\frac{6 N_f}{2\pi^2}\,\int_0^{p_F} dp\,p^2 \sqrt{p^2 + m^2 (\varepsilon)},
\end{equation}

where $N_f$ =3 is the number of flavors allowed for, the Fermi 
momentum $p_F\,=\,(\frac{\pi^2}{N_f \langle v\rangle})^{1/3}$, and closely
interrelated with each other vacuum current energy density and valonic masses are taken 
as follows:
        
\begin{equation}
\varepsilon_{vac}\,\equiv\,-P_{vac}
\simeq\,\varepsilon^0_{vac}\,\exp[-a\,(\varepsilon/|\varepsilon^0_{vac}|\,-\,1)]
\end{equation} 

and 

\begin{equation}
m_{u,d}\,\simeq\,m_0\,\exp[-a\,(\varepsilon/|\varepsilon^0_{vac}|\,-\,1)],
\end{equation}  

where $m_0\,\simeq\,\frac{1}{3}m_n\,\simeq$ 330 MeV \footnote{Actually, the 
numerical solution of eq.(1) allowed for the $\sim$150-MeV mass difference
between $(u,d)-$ and $s$-valons, but no significant 
correction was shown to come therefrom \cite{R_2009}.} 
and \,$a\,\sim$\,1 or larger 
is a free parameter, which describes the rate of QCD vacuum 
condensate destruction 
The numerical solution of eq.(1), supplemented with eq's.(2,3) is presented 
in Fig.2. Note, that only values $a\,\geq$\,1 are 
physically reasonable 
because the HPh vacuum condensate should be crucially affected by the 
particle energy density, as the latter one approaches the absolute value of the 
condensate strength itself (or even earlier). Below, in Fig.2, the curves 
2-4, which refer to $a\,<$\,1, are depicted for an illustration only. It is 
evident that hard scenario comes back in the limit $a\,\to\,\infty$.

\newpage

 \begin{figure}[h]
\includegraphics[width=10cm]{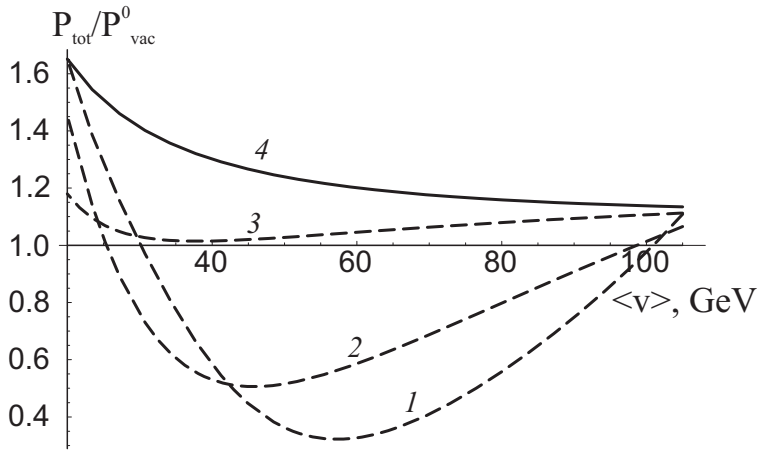}
\caption{The soft-scenario total pressure within the 
  SHPh-medium, $P_{tot}=P_{vac}+P_{part}$, as a function of particle specific 
  volume $\langle v\rangle$ at
  $\langle v\rangle\,\leq$ 100 GeV$^{-3}$,
  if the nuclear matter were "cold" (curves 1,2,3,4 refer to 
  $a$ = 1,\,\,0.5,\,\,0.1,\,\,0.01, respectively). It is evident that no 
  "cold" steady state of a star with quark center is accessible at    
  $a\,\geq$ 0.1, since the inequality $dP_{tot}/d\langle v\rangle\,>$ 0 
  signaling 
  of instability holds within some density interval to be passed along that way.}
\end{figure}

\vspace*{1cm}     

{Thus, we come to to the principally significant conclusion, that {\it no way 
exists for preserving degeneracy under HPh $\longrightarrow$ SHPh phase 
transition}.  

Does it rule out the possibility of BH formation in course of a 
compact star evolution? We try to put forward some seemingly weighty arguments 
that, indeed, it does.   

\newpage
 
\subsection{NS \, {\it vs} \, BH}


1.\quad {\bf The upper bound for NS}

As being emerged, the central domain of   
SHPh starts swelling until a balance is established between further heating 
due to gravitational compression and energy outflow. If the 
equilibrium SHPh mass is sufficiently large for making the real 
high-temperature quark-gluon plasma (QGP - nearly perfect gas of multiply 
produced qluons and
$q\bar q$-pairs, baryonic chemical potential thus becoming about zero) and yet
is small as compared to the total star mass, then a
reasonable (although elementary and crude) condition for the hydrodynamic
(fast process) equilibrium reads:

\begin{equation}
\frac{\pi^2}{30}(2\,\times\,8\,+
2\,\times\,3\,\times\,2\,\times\,3\,\times\,\frac{7}{8})\,T^4\,\,\simeq
\end{equation}
$$\frac{6}{7}\,G\,(\varepsilon_n + 3P_n + 2|\varepsilon^0_{vac}|)^2 \,V/R\,
\simeq\,
\frac{4\pi}{3}\,\frac{6}{7}\,(3\div 4)^2 G\,\varepsilon_n^2 R^2,$$\\

where the mean energy density of QGP at the star central interior (the left-hand 
side) is equated to that of HPh non-relativistic star main body. Here 
$T$ is the QGP temperature, $G$ is the gravitational constant, $R$ and $V$ stand 
for the star radius and volume, respectively. Also the "weigh of pressure" is 
taken into account, what is especially significant for the HPh vacuum:
$\varepsilon^0_{vac} + 3P^0_{vac}\,=\,2|\varepsilon^0_{vac}|\,\simeq\,
2\varepsilon_n$. After insertion in eq.(4) the proper numerical values, one 
obtains
                              
\begin{equation}  
T\,\simeq\,(170-200)\,\sqrt{k}\,
(\mbox{MeV}),
\end{equation}

where $k\,=\,R/10$ km. Since, according to the well known lattice simulations
\cite{Karsch}, QGP is expected to come into being just at  
$T\,\simeq\,(170-200)$ MeV, 
the hydrodynamic equilibrium between the first appeared hot SHPh at the star 
interior and cold HPh at its periphery could be maintained at $R\,\simeq$ 10 km 
and corresponding ("critical") star mass $M_{NS}\,\simeq\,2.3\,M_{\bigodot}$\footnote{
This estimate is in quite good agreement with the large body of data 
on the NS masses.},\,\,{\it but this equilibrium is achieved at the price of an enormous 
thermodynamic (slow process) disbalance} (note again that 
neutrinos get stuck at the relevant densities of nuclear matter). Thus, at
$M_{NS}\,\geq\,2.3\,M_{\bigodot}$, the thermal instability grows up resulting
in heat outflow, which gets more and more powerful along with $M_{NS}$ increase
and, hence, produces eruption of mass and energy which should result in 
the following observable phenomena:

--\,\, either in the successive GRB's (the more destructive 
ones the larger is $M_{NS}$, up to being ($10^4-10^5$) 
times as powerful as those emerged under the 
typical supernovae explosions, because the relevant temperatures differ by
more then one order), which stop as   
$M_{NS}$ grows down to become below the critical value, 
$M_{NS}\,\simeq\,2.3\,M_{\bigodot}$, since then no QGP forms at the star center;
          
--\,\, or in the total self-destruction of the star. \\

However, still one way of star evolution is conceivable:

--\,\,{\it  BH may shut to and trap the matter before the above mechanisms 
come into play}\\ \\ \\ 


2. \quad {\bf The lower bounds for BH rule it out ?}\\ \\

{The elementary condition for horizon first appearance within the body of a 
compact star reads: $\frac{2GM_g}{R_g}$ = 1, or

\begin{equation}
R_g\,\simeq\,\lbrack\frac{3}{8\pi 
G \langle\varepsilon_g \rangle }\rbrack^{1/2,}
\end{equation}

where $R_G$ and $\langle\varepsilon_g \rangle$ are the BH radius and its mean energy 
density, respectively. For getting the lower estimate of $R_G$, one 
has to take into account that $\langle\varepsilon_g \rangle\,\leq\,\varepsilon_n$ 
because the energy density profile maximizes at the star center but should,
nevertheless, not exceed the value $\varepsilon_n$ there (to escape
the premature phase transition instability there and all the cataclysms what follow, see 
above). Thus, we obtain from eq.(6)\\    

\centerline{$R_g\,\geq\,12$ km and $M_g\,\geq\,4\,M_{\bigodot}$}  
\vspace*{3mm}
and, therefore, {\it the strong NS instability (at $M_{NS}\,\geq\,2.3\,M_{\bigodot}$, see 
above) is expected to develop well before BH appearance.} 

One can by no means diminish the above BH 
parameters even allowing for powerful confluent 
shock waves caused by preceding supernovae 
explosion, because any attempt of such a kind would ask 
unavoidably for $\langle\varepsilon \rangle\,\geq\,\varepsilon_n$
inside the wave body itself, 
what, in turn, would result in the immediate 
developing of the aforementioned HPh$\longrightarrow$SHPh phase 
transition instability, enormous heating and, finally, 
in rupturing the shock wave body from within.}


\section{Conclusion}
 Two QCD-motivated "alarming" signals are put forward:    

\noindent --\,\, neutron stars of highest masses are in face of instability 
associated with QCD-vacuum transformation under HPh $\longrightarrow$ SHPh  
transition; 

\noindent --\,\, this instability makes rather problematic the accessibility
of a black hole configuration as the final state of collapsing compact star;\\

\noindent $\bullet$ \,\, It is difficult to resist the temptation of 
linking the instability under discussion and the poorly understood
data on very distant (young) GRB's of highest energy, 
like GRB 090423 \cite{krimm}, GRB 080916C \cite{abdo}, GRB 080319B  
("naked eye") \cite{bloom}, etc.      

The work is supported by RFBR, grant \#08-02-13637, and by RF President 
Foundation for The Leading Scientific Scools, grant \#NSH-438.2008.2.

\end{document}